\documentclass[%
 aip, pof,  
 amsmath,amssymb,
  reprint,  
]{revtex4-1}

\usepackage{graphicx}
\usepackage{dcolumn}
\usepackage{bm}

\begin{document}
 

\title[compressible turbulence inertial range]{Analytic derivation of the  inertial range   of compressible turbulence} 
\author{I. Goldman}
 \email{goldman@afeka.ac.il}
\affiliation{ 
Physics Department Afeka Engineering College, Tel Aviv, Israel
}%
\affiliation{Astrophysics Department, Tel Aviv University, Tel Aviv, Israel }

\date{\today}

\begin{abstract}
An analytic model for steady state turbulence is employed to obtain the inertial range power spectrum  of compressible turbulence.  
 We assume that for homogeneous turbulence, the timescales controlling the energy injected at a given  wavenumber from all smaller wave-numbers,   are equal for each spatial component.  However, the longitudinal component  energy is   diverted into compression, so the rate controlling the energy that is transferred to all larger wave-numbers by the turbulent viscosity is reduced. The resulting   inertial range  is a power law with index -2. Indeed such power  spectra were observed in various astrophysical settings and also  in numerical simulations.

\end{abstract}
 
\maketitle

Compressible supersonic turbulence characterized by  a 1D   power-law index of -2 has been observed in molecular clouds  \cite{Larson81, Leung+82, Dame+86},
in  interstellar medium of galaxies  \cite{Roy+Joncas85, Green93, Stanimirovic+99, Goldman2000}, in a shocked  nebula near the Galactic center \cite{contini+goldman2011}. It has also been obtained  
 in numerical simulations  \cite{Passot+88, V-Semadeni+97}. Recently, it was found \cite{Besserglick+Goldman2021}   that the intensity of $\gamma$-Ray emission from the Large Magellanic Cloud  exhibits such a   power spectrum.  
  
In this Letter, we employ a simple analytic  model  for steady  turbulence  \cite{Canuto+87} to find  the power spectrum of the inertial range of  supersonic  compressible turbulence. We obtain an inertial range  of the 1D power spectrum  which   is proportional to  $k^{-2}$. Details follow.
   
 The model  \cite{Canuto+87} is an extension of a previous  work \cite{Canuto+Goldman85}. It describes   homogeneous isotropic steady state turbulence.
 Its basic equations are
 
 \begin{eqnarray}
     n_s(k) + \frac{y(k)}{n_c(k)}= \nu_t (k) k^2 \\ 
 y(k)= \int_{k_0}^k F(k) k^2 dk\\ 
 \nu_t(k)= \int_k^{\infty} \frac{F(K)}{n_c(k)}\\
  \nu_t (k) k^2= \gamma n_c(k)
    \end{eqnarray}
    with $F(k)= 4 \pi k^2\Phi(k)$ where $k =|\vec k|$, and  $\Phi(\vec k)$ is the  3D power spectrum, of the turbulent velocity, $y(k)$ is the  k-space mean square  vorticity at wavenumber $k$, $n_s(k)$ is the rate controlling the energy input from the source at $k$, $n_c(k)$ is the inverse of the eddy correlation timescale
 and    $\nu_t(k)$ is the turbulent kinematic viscosity at wavenumber $k$ exerted by 
 all the eddies with larger wavenumbers. Eq.(4) relates the turbulent viscosity and the eddy correlation rate  , with $\gamma$ being a dimensionless   constant.
The model has been quite successful in obtaining turbulence  spectra  for both the large scale spatial scales as well  as the inertial range. An  extension of the model \cite{Canuto+96}dealing with stellar turbulent convection was  very successful and has been widely cited.

   In the large-eddy range $n_s(k)$ is positive and dominates over $\frac{y(k)}{nc(k)}$. In the dissipation range (small spatial scales) $n_s(k)= - \nu k^2$, with $\nu$denoting the microscopic kinematic viscosity. Depending on $n_s(k)$ and $\nu$, there usually  exists  a mid wavenumber range-- the inertial range, in which both energy  input from the source and energy  dissipation are very small compared to the rate of energy transfer. Thus, in this range
   \begin{equation}
 y(k)  =  \gamma n_c(k)^2
   \end{equation}
  
   from which 
   \begin{equation}
   y'(k)  =2  \gamma n_c(k) n_c'(k)
   \end{equation}
   
   From  Eq.(2) Eq.(3) and Eq.(4) follows that 
   
  \begin{eqnarray}
   y'(k) = - k^2 \nu_t '(k)= - \gamma k^2\left(n_c(k) k^{-2}\right)'
 \end{eqnarray}

  Eqs.(6) and (7)  yield 
   \begin{equation}
   n_c(k)= A k^{2/3}
   \end{equation}
with $A$ a constant. Since $F(k)= y'(k) k^{-2}$ one gets

\begin{equation}
F(k)=\frac{4}{3} \gamma A^2 k^{-5/3}
\end{equation}
Thus, the  model yields for the inertial range a power spectrum of the Kolmogorov \cite{Kolmogorov41} form . This is not surprising as in this model all the energy injected from smaller wave-numbers is transferred by the turbulent viscosity to the larger wave-numbers.

 However, in compressible turbulence the situation is different: not all the energy  available from the wave-numbers smaller than $k$ is cascaded to smaller scales, since energy is also being diverted  into compression. The longitudinal component of the turbulence is the one involved in the energy diversion. For an homogeneous turbulence it is conceivable that the timescales controlling the  input from the wave-numbers smaller than $k$ are equal. Thus, if all its energy is diverted to compression one gets    instead of Eq.(1) 
  \begin{equation}
        n_s(k) +\frac{2}{3} \frac{y(k)}{n_c(k)}= \nu_t (k) k^2= \gamma n_c(k)
      \end{equation} 
 The assumption that all the energy of the longitudinal component is diverted into compression, seems  justified for supesonic turbulence, which is ubiquitous in astrophysical turbulence. 
From Eq.(10) follows the rate equation   the inertial range:
  
   \begin{equation}
   y(k) =\frac{3}{2}\gamma n_c(k)^2 
  \end{equation}
  
  Eq.(7) is unchanged while differentiation of Eq.(11) yields
  
 \begin{equation}
   y'(k)  =3  \gamma n_c(k) n_c'(k)
   \end{equation}
  Thus, the solution is now

\begin{equation}
n_c(k)=B k^{1/2} 
\end{equation} 
with $B$  a constant, 
leading to  
 \begin{equation}
 F(k) =\frac{3}{2} \gamma B^2  k^{-2}
 \end{equation}
Which is the expected form of the inertial range of compressible turbulence.  The present derivation suggests that for sonic
turbulence not all of the longitudinal energy will be diverted. In this case, one may expect that the inertial range logarithmic slope will be intermediate between $-5/3$ and $-2$.

\section{data  availability}
  This is a theoretical paper. It includes no data.
 
\begin{acknowledgments} 
I thank of the Research Authority of Afeka College.
\end{acknowledgments}

\section*{references}
\nocite{*}
 \bibliography{comp_turb}

\providecommand{\noopsort}[1]{}\providecommand{\singleletter}[1]{#1}%
\begin{thebibliography}{15}%
\makeatletter
\providecommand \@ifxundefined [1]{%
 \@ifx{#1\undefined}
}%
\providecommand \@ifnum [1]{%
 \ifnum #1\expandafter \@firstoftwo
 \else \expandafter \@secondoftwo
 \fi
}%
\providecommand \@ifx [1]{%
 \ifx #1\expandafter \@firstoftwo
 \else \expandafter \@secondoftwo
 \fi
}%
\providecommand \natexlab [1]{#1}%
\providecommand \enquote  [1]{``#1''}%
\providecommand \bibnamefont  [1]{#1}%
\providecommand \bibfnamefont [1]{#1}%
\providecommand \citenamefont [1]{#1}%
\providecommand \href@noop [0]{\@secondoftwo}%
\providecommand \href [0]{\begingroup \@sanitize@url \@href}%
\providecommand \@href[1]{\@@startlink{#1}\@@href}%
\providecommand \@@href[1]{\endgroup#1\@@endlink}%
\providecommand \@sanitize@url [0]{\catcode `\\12\catcode `\$12\catcode
  `\&12\catcode `\#12\catcode `\^12\catcode `\_12\catcode `\%12\relax}%
\providecommand \@@startlink[1]{}%
\providecommand \@@endlink[0]{}%
\providecommand \url  [0]{\begingroup\@sanitize@url \@url }%
\providecommand \@url [1]{\endgroup\@href {#1}{\urlprefix }}%
\providecommand \urlprefix  [0]{URL }%
\providecommand \Eprint [0]{\href }%
\providecommand \doibase [0]{http://dx.doi.org/}%
\providecommand \selectlanguage [0]{\@gobble}%
\providecommand \bibinfo  [0]{\@secondoftwo}%
\providecommand \bibfield  [0]{\@secondoftwo}%
\providecommand \translation [1]{[#1]}%
\providecommand \BibitemOpen [0]{}%
\providecommand \bibitemStop [0]{}%
\providecommand \bibitemNoStop [0]{.\EOS\space}%
\providecommand \EOS [0]{\spacefactor3000\relax}%
\providecommand \BibitemShut  [1]{\csname bibitem#1\endcsname}%
\let\auto@bib@innerbib\@empty
\bibitem [{\citenamefont {{Larson}}(1981)}]{Larson81}%
  \BibitemOpen
  \bibfield  {author} {\bibinfo {author} {\bibfnamefont {R.~B.}\ \bibnamefont
  {{Larson}}},\ }\bibfield  {title} {\enquote {\bibinfo {title} {{Turbulence
  and star formation in molecular clouds.}}}\ }\href {\doibase
  10.1093/mnras/194.4.809} {\bibfield  {journal} {\bibinfo  {journal} {Monthly
  Notices of the Royal Astronomical Society}\ }\textbf {\bibinfo {volume}
  {194}},\ \bibinfo {pages} {809--826} (\bibinfo {year} {1981})}\BibitemShut
  {NoStop}%
\bibitem [{\citenamefont {{Leung}}, \citenamefont {{Kutner}},\ and\
  \citenamefont {{Mead}}(1982)}]{Leung+82}%
  \BibitemOpen
  \bibfield  {author} {\bibinfo {author} {\bibfnamefont {C.~M.}\ \bibnamefont
  {{Leung}}}, \bibinfo {author} {\bibfnamefont {M.~L.}\ \bibnamefont
  {{Kutner}}}, \ and\ \bibinfo {author} {\bibfnamefont {K.~N.}\ \bibnamefont
  {{Mead}}},\ }\bibfield  {title} {\enquote {\bibinfo {title} {{On the origin
  and structure of isolated dark globules.}}}\ }\href {\doibase 10.1086/160450}
  {\bibfield  {journal} {\bibinfo  {journal} {Astrophys. J.}\ }\textbf
  {\bibinfo {volume} {262}},\ \bibinfo {pages} {583--589} (\bibinfo {year}
  {1982})}\BibitemShut {NoStop}%
\bibitem [{\citenamefont {{Dame}}\ \emph {et~al.}(1986)\citenamefont {{Dame}},
  \citenamefont {{Elmegreen}}, \citenamefont {{Cohen}},\ and\ \citenamefont
  {{Thaddeus}}}]{Dame+86}%
  \BibitemOpen
  \bibfield  {author} {\bibinfo {author} {\bibfnamefont {T.~M.}\ \bibnamefont
  {{Dame}}}, \bibinfo {author} {\bibfnamefont {B.~G.}\ \bibnamefont
  {{Elmegreen}}}, \bibinfo {author} {\bibfnamefont {R.~S.}\ \bibnamefont
  {{Cohen}}}, \ and\ \bibinfo {author} {\bibfnamefont {P.}~\bibnamefont
  {{Thaddeus}}},\ }\bibfield  {title} {\enquote {\bibinfo {title} {{The Largest
  Molecular Cloud Complexes in the First Galactic Quadrant}},}\ }\href
  {\doibase 10.1086/164304} {\bibfield  {journal} {\bibinfo  {journal}
  {Astrophys. J.}\ }\textbf {\bibinfo {volume} {305}},\ \bibinfo {pages} {892}
  (\bibinfo {year} {1986})}\BibitemShut {NoStop}%
\bibitem [{\citenamefont {{Roy}}\ and\ \citenamefont
  {{Joncas}}(1985)}]{Roy+Joncas85}%
  \BibitemOpen
  \bibfield  {author} {\bibinfo {author} {\bibfnamefont {J.~R.}\ \bibnamefont
  {{Roy}}}\ and\ \bibinfo {author} {\bibfnamefont {G.}~\bibnamefont
  {{Joncas}}},\ }\bibfield  {title} {\enquote {\bibinfo {title} {{Structure and
  origin of velocity fluctuations in the HII region Sharpless 142.}}}\ }\href
  {\doibase 10.1086/162772} {\bibfield  {journal} {\bibinfo  {journal}
  {Astrophys. J.}\ }\textbf {\bibinfo {volume} {288}},\ \bibinfo {pages}
  {142--147} (\bibinfo {year} {1985})}\BibitemShut {NoStop}%
\bibitem [{\citenamefont {{Green}}(1993)}]{Green93}%
  \BibitemOpen
  \bibfield  {author} {\bibinfo {author} {\bibfnamefont {D.~A.}\ \bibnamefont
  {{Green}}},\ }\href {\doibase 10.1093/mnras/262.2.327} {\enquote {\bibinfo
  {title} {{A power spectrum analysis of the angular scale of Galactic neutral
  hydrogen emission towards L = 140 deg, B = 0 deg}},}\ } (\bibinfo {year}
  {1993})\BibitemShut {NoStop}%
\bibitem [{\citenamefont {{Stanimirovic}}\ \emph {et~al.}(1999)\citenamefont
  {{Stanimirovic}}, \citenamefont {{Staveley-Smith}}, \citenamefont {{Dickey}},
  \citenamefont {{Sault}},\ and\ \citenamefont {{Snowden}}}]{Stanimirovic+99}%
  \BibitemOpen
  \bibfield  {author} {\bibinfo {author} {\bibfnamefont {S.}~\bibnamefont
  {{Stanimirovic}}}, \bibinfo {author} {\bibfnamefont {L.}~\bibnamefont
  {{Staveley-Smith}}}, \bibinfo {author} {\bibfnamefont {J.~M.}\ \bibnamefont
  {{Dickey}}}, \bibinfo {author} {\bibfnamefont {R.~J.}\ \bibnamefont
  {{Sault}}}, \ and\ \bibinfo {author} {\bibfnamefont {S.~L.}\ \bibnamefont
  {{Snowden}}},\ }\bibfield  {title} {\enquote {\bibinfo {title} {{The
  large-scale HI structure of the Small Magellanic Cloud}},}\ }\href {\doibase
  10.1046/j.1365-8711.1999.02013.x} {\bibfield  {journal} {\bibinfo  {journal}
  {Monthly Notices of the Royal Astronomical Society}\ }\textbf {\bibinfo
  {volume} {302}},\ \bibinfo {pages} {417--436} (\bibinfo {year}
  {1999})}\BibitemShut {NoStop}%
\bibitem [{\citenamefont {{Goldman}}(2000)}]{Goldman2000}%
  \BibitemOpen
  \bibfield  {author} {\bibinfo {author} {\bibfnamefont {I.}~\bibnamefont
  {{Goldman}}},\ }\bibfield  {title} {\enquote {\bibinfo {title}
  {{Interpretation of the Spatial Power Spectra of Neutral Hydrogen in the
  Galaxy and in the Small Magellanic Cloud}},}\ }\href {\doibase
  10.1086/309456} {\bibfield  {journal} {\bibinfo  {journal} {Astrophys. J.}\
  }\textbf {\bibinfo {volume} {541}},\ \bibinfo {pages} {701--706} (\bibinfo
  {year} {2000})},\ \Eprint {http://arxiv.org/abs/astro-ph/0005023}
  {arXiv:astro-ph/0005023 [astro-ph]} \BibitemShut {NoStop}%
\bibitem [{\citenamefont {{Contini}}\ and\ \citenamefont
  {{Goldman}}(2011)}]{contini+goldman2011}%
  \BibitemOpen
  \bibfield  {author} {\bibinfo {author} {\bibfnamefont {M.}~\bibnamefont
  {{Contini}}}\ and\ \bibinfo {author} {\bibfnamefont {I.}~\bibnamefont
  {{Goldman}}},\ }\bibfield  {title} {\enquote {\bibinfo {title} {{Spectra from
  the shocked nebulae revealing turbulence near the Galactic Centre}},}\ }\href
  {\doibase 10.1111/j.1365-2966.2010.17719.x} {\bibfield  {journal} {\bibinfo
  {journal} {Monthly Notices of the Royal Astronomical Society}\ }\textbf
  {\bibinfo {volume} {411}},\ \bibinfo {pages} {792--806} (\bibinfo {year}
  {2011})}\BibitemShut {NoStop}%
\bibitem [{\citenamefont {{Passot}}, \citenamefont {{Pouquet}},\ and\
  \citenamefont {{Woodward}}(1988)}]{Passot+88}%
  \BibitemOpen
  \bibfield  {author} {\bibinfo {author} {\bibfnamefont {T.}~\bibnamefont
  {{Passot}}}, \bibinfo {author} {\bibfnamefont {A.}~\bibnamefont {{Pouquet}}},
  \ and\ \bibinfo {author} {\bibfnamefont {P.}~\bibnamefont {{Woodward}}},\
  }\bibfield  {title} {\enquote {\bibinfo {title} {{The plausibility of
  Kolmogorov-type spectra in molecular clouds}},}\ }\href@noop {} {\bibfield
  {journal} {\bibinfo  {journal} {Astronomy \& Astrophysics}\ }\textbf
  {\bibinfo {volume} {197}},\ \bibinfo {pages} {228--234} (\bibinfo {year}
  {1988})}\BibitemShut {NoStop}%
\bibitem [{\citenamefont {{V{\'a}zquez-Semadeni}}, \citenamefont
  {{Ballesteros-Paredes}},\ and\ \citenamefont
  {{Rodr{\'\i}guez}}(1997)}]{V-Semadeni+97}%
  \BibitemOpen
  \bibfield  {author} {\bibinfo {author} {\bibfnamefont {E.}~\bibnamefont
  {{V{\'a}zquez-Semadeni}}}, \bibinfo {author} {\bibfnamefont {J.}~\bibnamefont
  {{Ballesteros-Paredes}}}, \ and\ \bibinfo {author} {\bibfnamefont {L.~F.}\
  \bibnamefont {{Rodr{\'\i}guez}}},\ }\bibfield  {title} {\enquote {\bibinfo
  {title} {{A Search for Larson-type Relations in Numerical Simulations of the
  ISM: Evidence for Nonconstant Column Densities}},}\ }\href {\doibase
  10.1086/303432} {\bibfield  {journal} {\bibinfo  {journal} {Astrophys. J.}\
  }\textbf {\bibinfo {volume} {474}},\ \bibinfo {pages} {292--307} (\bibinfo
  {year} {1997})},\ \Eprint {http://arxiv.org/abs/astro-ph/9607175}
  {arXiv:astro-ph/9607175 [astro-ph]} \BibitemShut {NoStop}%
\bibitem [{\citenamefont {{Besserglik}}\ and\ \citenamefont
  {{Goldman}}(2021)}]{Besserglick+Goldman2021}%
  \BibitemOpen
  \bibfield  {author} {\bibinfo {author} {\bibfnamefont {D.}~\bibnamefont
  {{Besserglik}}}\ and\ \bibinfo {author} {\bibfnamefont {I.}~\bibnamefont
  {{Goldman}}},\ }\bibfield  {title} {\enquote {\bibinfo {title} {{The power
  spectrum and structure function of the Gamma Ray emission from the Large
  Magellanic Cloud}},}\ }\href@noop {} {\bibfield  {journal} {\bibinfo
  {journal} {arXiv e-prints}\ ,\ \bibinfo {eid} {arXiv:2103.09531}} (\bibinfo
  {year} {2021})},\ \Eprint {http://arxiv.org/abs/2103.09531} {arXiv:2103.09531
  [astro-ph.GA]} \BibitemShut {NoStop}%
\bibitem [{\citenamefont {{Canuto}}, \citenamefont {{Goldman}},\ and\
  \citenamefont {{Chasnov}}(1987)}]{Canuto+87}%
  \BibitemOpen
  \bibfield  {author} {\bibinfo {author} {\bibfnamefont {V.~M.}\ \bibnamefont
  {{Canuto}}}, \bibinfo {author} {\bibfnamefont {I.}~\bibnamefont {{Goldman}}},
  \ and\ \bibinfo {author} {\bibfnamefont {J.}~\bibnamefont {{Chasnov}}},\
  }\bibfield  {title} {\enquote {\bibinfo {title} {{A model for fully developed
  turbulence}},}\ }\href {\doibase 10.1063/1.866472} {\bibfield  {journal}
  {\bibinfo  {journal} {Physics of Fluids}\ }\textbf {\bibinfo {volume} {30}},\
  \bibinfo {pages} {3391--3418} (\bibinfo {year} {1987})}\BibitemShut {NoStop}%
\bibitem [{\citenamefont {{Canuto}}\ and\ \citenamefont
  {{Goldman}}(1985)}]{Canuto+Goldman85}%
  \BibitemOpen
  \bibfield  {author} {\bibinfo {author} {\bibfnamefont {V.~M.}\ \bibnamefont
  {{Canuto}}}\ and\ \bibinfo {author} {\bibfnamefont {I.}~\bibnamefont
  {{Goldman}}},\ }\bibfield  {title} {\enquote {\bibinfo {title} {{Analytical
  model for large-scale turbulence}},}\ }\href {\doibase
  10.1103/PhysRevLett.54.430} {\bibfield  {journal} {\bibinfo  {journal}
  {PhysRevLett}\ }\textbf {\bibinfo {volume} {54}},\ \bibinfo {pages}
  {430--433} (\bibinfo {year} {1985})}\BibitemShut {NoStop}%
\bibitem [{\citenamefont {{Canuto}}, \citenamefont {{Goldman}},\ and\
  \citenamefont {{Mazzitelli}}(1996)}]{Canuto+96}%
  \BibitemOpen
  \bibfield  {author} {\bibinfo {author} {\bibfnamefont {V.~M.}\ \bibnamefont
  {{Canuto}}}, \bibinfo {author} {\bibfnamefont {I.}~\bibnamefont {{Goldman}}},
  \ and\ \bibinfo {author} {\bibfnamefont {I.}~\bibnamefont {{Mazzitelli}}},\
  }\bibfield  {title} {\enquote {\bibinfo {title} {{Stellar Turbulent
  Convection: A Self-consistent Model}},}\ }\href {\doibase 10.1086/178166}
  {\bibfield  {journal} {\bibinfo  {journal} {Astrophys. J.}\ }\textbf
  {\bibinfo {volume} {473}},\ \bibinfo {pages} {550} (\bibinfo {year}
  {1996})},\ \Eprint {http://arxiv.org/abs/astro-ph/9609001}
  {arXiv:astro-ph/9609001 [astro-ph]} \BibitemShut {NoStop}%
\bibitem [{\citenamefont {{Kolmogorov}}(1941)}]{Kolmogorov41}%
  \BibitemOpen
  \bibfield  {author} {\bibinfo {author} {\bibfnamefont {A.}~\bibnamefont
  {{Kolmogorov}}},\ }\bibfield  {title} {\enquote {\bibinfo {title} {{The Local
  Structure of Turbulence in Incompressible Viscous Fluid for Very Large
  Reynolds' Numbers}},}\ }\href@noop {} {\bibfield  {journal} {\bibinfo
  {journal} {Akademiia Nauk SSSR Doklady}\ }\textbf {\bibinfo {volume} {30}},\
  \bibinfo {pages} {301--305} (\bibinfo {year} {1941})}\BibitemShut {NoStop}%
\end{thebibliography}%
 
  \end{document}